\documentclass[ALICE,manyauthors]{cernphprep}
\usepackage[comma,square,numbers,sort&compress]{natbib}
\usepackage{hyperref}
\usepackage{lineno}
\usepackage{xspace}
\usepackage{adjustbox}
\usepackage{rotating}  
\usepackage{booktabs}
\usepackage{xcolor}
\usepackage[T1]{fontenc}
\usepackage{orcidlink}

\begin{document}

\newcommand{\pp}           {pp\xspace}
\newcommand{\ppbar}        {\mbox{$\mathrm {p\overline{p}}$}\xspace}
\newcommand{\XeXe}         {\mbox{Xe--Xe}\xspace}
\newcommand{\PbPb}         {\mbox{Pb--Pb}\xspace}
\newcommand{\pA}           {\mbox{pA}\xspace}
\newcommand{\pPb}          {\mbox{p--Pb}\xspace}
\newcommand{\AuAu}         {\mbox{Au--Au}\xspace}
\newcommand{\dAu}          {\mbox{d--Au}\xspace}

\newcommand{\s}            {\ensuremath{\sqrt{s}}\xspace}
\newcommand{\snn}          {\ensuremath{\sqrt{s_{\mathrm{NN}}}}\xspace}
\newcommand{\pt}           {\ensuremath{p_{\rm T}}\xspace}
\newcommand{\meanpt}       {$\langle p_{\mathrm{T}}\rangle$\xspace}
\newcommand{\ycms}         {\ensuremath{y_{\rm CMS}}\xspace}
\newcommand{\ylab}         {\ensuremath{y_{\rm lab}}\xspace}
\newcommand{\etarange}[1]  {\mbox{$\left | \eta \right |~<~#1$}}
\newcommand{\yrange}[1]    {\mbox{$\left | y \right |~<~#1$}}
\newcommand{\dndy}         {\ensuremath{\mathrm{d}N_\mathrm{ch}/\mathrm{d}y}\xspace}
\newcommand{\dndeta}       {\ensuremath{\mathrm{d}N_\mathrm{ch}/\mathrm{d}\eta}\xspace}
\newcommand{\avdndeta}     {\ensuremath{\langle\dndeta\rangle}\xspace}
\newcommand{\dNdy}         {\ensuremath{\mathrm{d}N_\mathrm{ch}/\mathrm{d}y}\xspace}
\newcommand{\Npart}        {\ensuremath{N_\mathrm{part}}\xspace}
\newcommand{\Ncoll}        {\ensuremath{N_\mathrm{coll}}\xspace}
\newcommand{\dEdx}         {\ensuremath{\textrm{d}E/\textrm{d}x}\xspace}
\newcommand{\RpPb}         {\ensuremath{R_{\rm pPb}}\xspace}

\newcommand{\nineH}        {$\sqrt{s}~=~0.9$~Te\kern-.1emV\xspace}
\newcommand{\seven}        {$\sqrt{s}~=~7$~Te\kern-.1emV\xspace}
\newcommand{\twoH}         {$\sqrt{s}~=~0.2$~Te\kern-.1emV\xspace}
\newcommand{\twosevensix}  {$\sqrt{s}~=~2.76$~Te\kern-.1emV\xspace}
\newcommand{\five}         {$\sqrt{s}~=~5.02$~Te\kern-.1emV\xspace}
\newcommand{\twosevensixnn}{$\sqrt{s_{\mathrm{NN}}}~=~2.76$~Te\kern-.1emV\xspace}
\newcommand{\fivenn}       {$\sqrt{s_{\mathrm{NN}}}~=~5.02$~Te\kern-.1emV\xspace}
\newcommand{\LT}           {L{\'e}vy-Tsallis\xspace}
\newcommand{\GeVc}         {Ge\kern-.1emV/$c$\xspace}
\newcommand{\MeVc}         {Me\kern-.1emV/$c$\xspace}
\newcommand{\TeV}          {Te\kern-.1emV\xspace}
\newcommand{\GeV}          {Ge\kern-.1emV\xspace}
\newcommand{\MeV}          {Me\kern-.1emV\xspace}
\newcommand{\GeVmass}      {Ge\kern-.2emV/$c^2$\xspace}
\newcommand{\MeVmass}      {Me\kern-.2emV/$c^2$\xspace}
\newcommand{\lumi}         {\ensuremath{\mathcal{L}}\xspace}

\newcommand{\ITS}          {\rm{ITS}\xspace}
\newcommand{\TOF}          {\rm{TOF}\xspace}
\newcommand{\ZDC}          {\rm{ZDC}\xspace}
\newcommand{\ZDCs}         {\rm{ZDCs}\xspace}
\newcommand{\ZNA}          {\rm{ZNA}\xspace}
\newcommand{\ZNC}          {\rm{ZNC}\xspace}
\newcommand{\SPD}          {\rm{SPD}\xspace}
\newcommand{\SDD}          {\rm{SDD}\xspace}
\newcommand{\SSD}          {\rm{SSD}\xspace}
\newcommand{\TPC}          {\rm{TPC}\xspace}
\newcommand{\TRD}          {\rm{TRD}\xspace}
\newcommand{\VZERO}        {\rm{V0}\xspace}
\newcommand{\VZEROA}       {\rm{V0A}\xspace}
\newcommand{\VZEROC}       {\rm{V0C}\xspace}
\newcommand{\Vdecay} 	   {\ensuremath{V^{0}}\xspace}

\newcommand{\ee}           {\ensuremath{e^{+}e^{-}}} 
\newcommand{\pip}          {\ensuremath{\pi^{+}}\xspace}
\newcommand{\pim}          {\ensuremath{\pi^{-}}\xspace}
\newcommand{\kap}          {\ensuremath{\rm{K}^{+}}\xspace}
\newcommand{\kam}          {\ensuremath{\rm{K}^{-}}\xspace}
\newcommand{\pbar}         {\ensuremath{\rm\overline{p}}\xspace}
\newcommand{\kzero}        {\ensuremath{{\rm K}^{0}_{\rm{S}}}\xspace}
\newcommand{\lmb}          {\ensuremath{\Lambda}\xspace}
\newcommand{\almb}         {\ensuremath{\overline{\Lambda}}\xspace}
\newcommand{\Om}           {\ensuremath{\Omega^-}\xspace}
\newcommand{\Mo}           {\ensuremath{\overline{\Omega}^+}\xspace}
\newcommand{\X}            {\ensuremath{\Xi^-}\xspace}
\newcommand{\Ix}           {\ensuremath{\overline{\Xi}^+}\xspace}
\newcommand{\Xis}          {\ensuremath{\Xi^{\pm}}\xspace}
\newcommand{\Oms}          {\ensuremath{\Omega^{\pm}}\xspace}
\newcommand{\degree}       {\ensuremath{^{\rm o}}\xspace}

\begin{titlepage}
  \PHyear{2026}
  \PHnumber{293}
  \PHdate{27 October}  
\title{Exploring the limit of partonic collectivity in small collision systems at the LHC}
\ShortTitle{Multiplicity dependent flow for identified hadrons}   
\Collaboration{ALICE Collaboration\footnote{See Appendix~\ref{app:collab} for the list of collaboration members}}
\ShortAuthor{ALICE Collaboration} 
\begin{abstract}
The existence of a partonic phase in small collision systems has been investigated previously through measurements of the elliptic flow ($v_{2}$) of mesons ($\mathrm{\pi}^{\pm}$, $\mathrm{K}^{\pm}$, and $\mathrm{K}_{S}^{0}$) and baryons ($\mathrm{p}(\overline{\mathrm{p}})$, $\Lambda(\overline{\Lambda})$) in high-multiplicity p--Pb collisions at $\sqrt{s_{\rm NN}}$~=~5.02~TeV and pp collisions at $\sqrt{s}$~=~13~TeV recorded by ALICE. The results have revealed a grouping and splitting behavior of $v_{2}$ for baryons and mesons at intermediate transverse momentum ($p_{\mathrm{T}}$), establishing that partonic collectivity, as observed in heavy-ion collisions, is also present in these systems. 
In this Letter, the $p_{\mathrm{T}}$-differential $v_{2}$ of various hadronic species is investigated in p--Pb collisions for the multiplicity intervals $N_{\rm ch} \geq 50$, $25 \leq N_{\rm ch} < 50$, and $15 \leq N_{\rm ch} < 25$, as well as in pp collisions for $25 \leq N_{\rm ch} < 50$ and $15 \leq N_{\rm ch} < 25$. Here, $N_{\rm ch}$ is the efficiency-corrected number of charged particles produced in the kinematic region of $|\eta|< 0.8$ and $0.2 < p_{\rm T} < 3.0$ GeV/$c$. The characteristic mass dependence of $v_{2}$, as well as the grouping and splitting patterns among identified particles, are examined.
The results in both p--Pb and pp collisions with $N_{\rm ch} \geq 25$ are consistent with the development of partonic collectivity. 
In the low-multiplicity region, $15 \leq N_{\rm ch} < 25$, the baryon-to-meson $v_{2}$ splitting, as quantified by the baryon-to-meson $v_2$ ratios, remains broadly consistent with that at higher multiplicities. However, its significance decreases to $\sim2\sigma$ in p--Pb collisions and to less than $1\sigma$ in pp collisions owing to the sizable experimental uncertainties.
These measurements constitute the first steps towards identifying the limits of partonic flow in small collision systems at the LHC.

\end{abstract}
\end{titlepage}

\setcounter{page}{2} 

\section{Introduction}
\label{sec:intro}
Ultrarelativistic heavy-ion collisions at the Super Proton Synchrotron (SPS), the Relativistic Heavy Ion Collider (RHIC), and the Large Hadron Collider (LHC) can produce a deconfined state of matter known as the quark--gluon plasma (QGP)~\cite{Shuryak:1978ij,ALICE:2022wpn,STAR:2005gfr,BRAHMS:2004adc,PHENIX:2004vcz,PHOBOS:2004zne,CMS:2024krd}. Evidence for the QGP formation includes phenomena such as jet quenching~\cite{ALICE:2021mqf,STAR:2020xiv,CMS:2012ulu,ATLAS:2022vii,NA50:2000brc} and $J/\psi$ suppression~\cite{Matsui:1986dk,NA50:2000mfb,PHENIX:2006gsi,ATLAS:2010xzb,ALICE:2012jsl}, while measurements of anisotropic flow provide important information on the properties and the collective dynamics of the produced system~\cite{Ollitrault:1997vz,ALICE:2011ab,ALICE:2016ccg,ALICE:2018lao,ALICE:2019zfl,STAR:2004jwm,STAR:2022gki,PHENIX:2011yyh,PHENIX:2018lia,CMS:2012zex,CMS:2013jlh,ALICE:2012vgf,ALICE:2014wao,ALICE:2018yph,ALICE:2022zks,STAR:2007afq,PHENIX:2006dpn}. Similar phenomena have also been investigated in small collision systems, such as proton--lead (p--Pb) and proton--proton (pp) collisions at the LHC, to assess the possible emergence of QGP-like matter~\cite{ALICE:2016fzo,ALICE:2024vzv}. Several investigations based on two-particle correlations in pseudorapidity ($\Delta\eta$) and azimuthal angle ($\Delta\varphi$) between particle pairs reported the observation of characteristic long-range correlation structures known as``ridges"~\cite{CMS:2010ifv,CMS:2012qk,ALICE:2011svq,CMS:2016fnw,ALICE:2013snk,ALICE:2023gyf,Abelev:2012ola,ATLAS:2013jmi}.
These long-range correlations, which are explained by anisotropic flow in large systems, motivate similar investigations in small collision systems. Anisotropic flow, which characterizes the azimuthal anisotropy in the transverse plane~\cite{Ollitrault:1992bk}, is quantified by the Fourier decomposition of the final-state particle azimuthal distribution, according to~\cite{Voloshin:1994mz}:
\begin{equation}
  \frac{\mathrm{d} N}{\mathrm{d} \varphi}=\frac{N}{2 \pi}\left[1+2 \sum_n v_n \cos \left[n\left(\varphi-\Psi_n\right)\right]\right], n=1,2,3, \ldots
\end{equation}
where $\varphi$ represents the azimuthal angle of the final-state particles.
The coefficients $v_{n}$ quantify the $n$-th harmonic flow, while $\Psi_{n}$ denotes the orientation of the corresponding symmetry plane. These harmonic flow coefficients characterize the momentum anisotropy of emitted particles and are sensitive both to the initial geometry of the collision system and to the subsequent dynamical evolution of the medium~\cite{PHENIX:2018lia,Alver:2010gr,Song:2017wtw,Kolb:2003dz}. The second-order coefficient $v_{2}$, known as elliptic flow, typically has the largest magnitude and arises in part from the almond-shaped overlap region in non-central heavy-ion collisions.
Measurements of centrality-dependent flow coefficients for inclusive hadrons in Pb--Pb and Xe--Xe collisions at the LHC, together with comparisons to hydrodynamic calculations, indicate that the QGP behaves as a nearly perfect fluid~\cite{STAR:2005gfr,ALICE:2011ab, ALICE:2016ccg, ALICE:2018lao, ALICE:2022wpn}. For the transverse-momentum-dependent elliptic flow, $v_{2}(p_{\rm T})$, radial expansion in an azimuthally anisotropic environment leads to a characteristic mass ordering of $v_{2}(p_{\rm T})$ in the low-momentum region ($p_{\rm T}<3$~GeV/$c$)~\cite{ALICE:2014wao,ALICE:2018yph,ALICE:2022zks,ALICE:2021ibz}. In the intermediate $p_{\mathrm{T}}$~region ($3<p_{\rm T}<6$~GeV/$c$), measurements show a grouping of $v_{2}$ for baryons and mesons, accompanied by a distinct splitting between the two hadron classes~\cite{,ALICE:2014wao,ALICE:2018yph,ALICE:2022zks,ALICE:2021ibz}. This behavior can be explained by the quark-coalescence mechanism~\cite{Sato:1981ez,Zhao:2020wcd,YuanyuanWang:2023vor,Fries:2008hs,Molnar:2003ff,Cao:2013ita}, in which hadrons are formed from partons in the hot and dense medium created in high-energy collisions, with their kinematic properties reflecting those of the constituent quarks. Consequently, the observed baryon--meson grouping and splitting is regarded as a signature of partonic collectivity in heavy-ion collisions~\cite{ALICE:2022wpn}.

In addition to studies performed in Pb--Pb and Xe--Xe collisions, investigations of anisotropic flow have also been carried out in small collision systems. The discovery of the ``ridge" structure in high-multiplicity pp and p--Pb collisions~\cite{CMS:2010ifv,ALICE:2013snk,ATLAS:2015hzw} motivated extensive studies of collective behavior in small collision systems. These investigations were subsequently extended to multiparticle azimuthal correlations~\cite{ALICE:2013snk,ATLAS:2015hzw,PHENIX:2018lia,CMS:2010ifv,CMS:2014und,CMS:2018loe,ALICE:2019zfl}, which provided evidence for long-range multiparticle collectivity in small systems. Measurements of the pseudorapidity dependence of anisotropic flow~\cite{ALICE:2023gyf} and of flow-vector fluctuations in the longitudinal direction~\cite{ALICE:2026suk} have further advanced the characterization of the three-dimensional initial geometry and its event-by-event fluctuations in small systems. Together with recent studies of correlations between anisotropic flow and mean transverse momentum~\cite{ALICE:2026cly}, which probe the interplay between the initial geometry and system size, measurements across collision systems at RHIC~\cite{PHENIX:2018lia} and the LHC suggest that anisotropic flow in small systems is also driven by these initial-state properties and their fluctuations. To further investigate the dynamical evolution of the medium created in small collision systems, ALICE recently reported measurements of $v_{2}(p_{\rm T})$ for mesons ($\mathrm{\pi}^{\pm}$, $\mathrm{K}^{\pm}$, and $\mathrm{K}_{S}^{0}$) and baryons ($\mathrm{p}(\overline{\mathrm{p}})$~and $\Lambda(\overline{\Lambda})$) in high-multiplicity p--Pb and pp collisions at the LHC. These events correspond to an average charged-particle multiplicity of $\langle N_{\rm ch} \rangle \approx 35$ in the kinematic region $|\eta|< 0.8$ and $0.2 < p_{\rm T} < 3.0$ GeV/$c$, after an event-by-event correction for tracking efficiency~\cite{ALICE:2024vzv}. The results show a clear mass ordering effect at low $p_{\mathrm{T}}$. In the intermediate $p_{\mathrm{T}}$~region, baryon--meson grouping (within deviations of less than $1\sigma$) and splitting (exceeding $5\sigma$ significance) of $v_{2}$ are observed. These results exhibit features analogous to those in heavy-ion collisions~\cite{ALICE:2012vgf,ALICE:2014wao,ALICE:2018yph,ALICE:2022zks,STAR:2007afq,PHENIX:2006dpn}. They can be described by model calculations incorporating the quark-coalescence hadronisation mechanism~\cite{Zhao:2020wcd,YuanyuanWang:2023vor}, whereas models lacking quark coalescence fail to reproduce the observed baryon--meson grouping of $v_{2}$. These results suggest the presence of partonic collectivity in small collision systems in the high-multiplicity region~\cite{ALICE:2024vzv}.
Furthermore, ultra-long-range correlations ($|\Delta \eta|>5.0$ and $6.5$) have been observed, visible down to $N_{\rm ch}\sim 8$ in p--Pb collisions and $N_{\rm ch}\sim 10$ in pp collisions at the LHC~\cite{ALICE:2025bwp}. 
When compared with theoretical model calculations, these ultra-long-range correlations in low-multiplicity p--Pb and pp collisions are attributed to final-state effects rather than initial-state effects such as the initial momentum anisotropy discussed in Refs.~\cite{Gelis:2010nm,BRAHMS:2004adc}. These final-state effects are driven by the initial geometry and its event-by-event fluctuations, which manifest through interactions among the produced particles at either the partonic or hadronic level~\cite{Zhou:2015iba,Bierlich:2024odg}. 
Despite the presence of ultra-long-range correlations in low-multiplicity events, it remains unclear whether this feature signifies partonic collectivity. Motivated by these observations, measurements of $v_{2}(p_{\rm T})$ for identified particles across various multiplicity regions enable the study of the evolution of azimuthal correlations in small collision systems. The multiplicity dependence of baryon--meson grouping and splitting patterns constitutes a key observable for assessing the extent to which signatures commonly associated with partonic collectivity persist toward low-multiplicity events.

This Letter reports on results of $v_{2}(p_{\rm T})$ for identified particles in p--Pb collisions at $\sqrt{s_{\rm NN}} = 5.02$~TeV and in pp collisions at $\sqrt{s} = 13$~TeV. The measurements are performed across multiple multiplicity intervals, including $N_{\rm ch} \geq 50$, $25 \leq N_{\rm ch} < 50$, and $15 \leq N_{\rm ch} < 25$. Furthermore, the ratios of $v_{2}(p_{\rm T})$ for baryons to mesons are calculated within these multiplicity intervals to quantitatively characterize the baryon--meson splitting pattern in both p--Pb and pp collisions.

\section{Experimental setup}
The p--Pb data at $\sqrt{s_{\rm NN}}=5.02$ TeV and pp data at $\sqrt{s}=13$ TeV used in this analysis were recorded with the ALICE detector during the LHC Run~2 data-taking period (2016--2018). The primary vertex (PV) of each collision event is reconstructed by combining data from the TPC and the Inner Tracking System (ITS)~\cite{Aamodt:2008zz}. The $z$-component of the primary vertex ($\text{PV}_{z}$) along the beam axis is required to be within $\pm 8$ cm to ensure uniform detector acceptance. In-bunch pileup is mitigated by rejecting events with more than one reconstructed primary vertex. Minimum-bias (MB) triggers based on signals from SPD and V0 detectors are used to suppress beam-gas and other machine-induced background events. For both pp and p--Pb collisions, the MB trigger relies on signals from the hybrid Silicon Pixel Detector (SPD) in the ITS and the V0 detector system. The V0 detector system~\cite{Abbas:2013taa}, located on either side of the central barrel, comprises arrays of scintillator tiles that cover the forward rapidity regions: $2.8 <\eta<5.1$ (V0A) and $-3.7<\eta<-1.7$ (V0C). For pp collisions, the MB trigger requires at least one hit in the SPD or in either V0 array. In p--Pb collisions, the MB trigger is defined by a coincidence of signals in both V0A and V0C. After applying selection criteria, the data samples contain approximately $198 \times 10^6$ events for p--Pb collisions and $226 \times 10^6$ events for pp collisions. The Forward Multiplicity Detectors (FMD)~\cite{Aamodt:2008zz,Abelev:2014ffa,ALICE:2004ftm} enable a large pseudorapidity gap between central and forward particles, which suppresses contamination from short-range correlations. The FMD system consists of three sub-detectors: FMD1 and FMD2 (collectively referred to as FMDA, spanning $1.7<\eta<5.0$), and FMD3 (FMDC, covering $-3.4<\eta<-1.7$). All FMD detectors utilize silicon strip sensor modules to measure the energy deposition of traversing particles. Out-of-bunch pileup events are excluded by correlating the FMD and V0 amplitudes, thereby reducing contamination from beam-induced background~\cite{ALICE:2023gyf}. The potential systematic effects of the event selection criteria are addressed in section~\ref{sec:syst}.

 For this analysis, tracks are reconstructed in the Time Projection Chamber (TPC)~\cite{Alme:2010ke} and are required to have at least 70 TPC space points (out of a maximum of 159) in the track fit and a $\chi^{2}_{\mathrm{TPC}}/\mathrm{NDF}$ of the TPC track fit less than 2.5, where NDF is the number of degrees of freedom. Consistency between the ITS and TPC track information is also required. To suppress secondary particle contamination, selected tracks are required to satisfy $\mathrm{DCA}_{z}<2$~cm, where $\mathrm{DCA}_{z}$ denotes the distance of closest approach (DCA) to the primary vertex along the beam axis. In addition, a $p_{\mathrm{T}}$-dependent requirement is imposed on the transverse DCA ($\mathrm{DCA}_{xy}$), where the requirement is given by $\mathrm{DCA}_{xy} < 7  \times(0.0026+0.005/p_{\mathrm{T}}^{1.01})~\mathrm{cm}$. Implementing such stringent DCA selection criteria significantly reduces contamination from particles produced by weak decays and interactions with the detector material, as demonstrated in~\cite{ALICE:2018vuu}. The tracking efficiency is calculated using Monte Carlo (MC) simulations and depends on transverse momentum and pseudorapidity as well as on the event multiplicity and the data-taking period. The integrated tracking efficiency of charged particles is approximately 78\% in p--Pb collisions and 83\% in pp collisions.

A Bayesian approach~\cite{ALICE:2016zzl,ALICE:2018yph} is used for particle-identification (PID), classifying particles as $\mathrm{\pi}^{\pm}$, $\mathrm{K}^{\pm}$, and $\mathrm{p}(\overline{\mathrm{p}})$~by combining information from the TPC and the Time-of-Flight (TOF) detector~\cite{Carnesecchi:2018oss}. Identification thresholds of 95\%, 90\%, and 85\% are set for $\mathrm{\pi}^{\pm}$, $\mathrm{K}^{\pm}$, and $\mathrm{p}(\overline{\mathrm{p}})$, respectively, resulting in purities above 95\%, 88\%, and 91\% as determined from MC simulations for p--Pb and pp collisions. $\mathrm{K}_{S}^{0}$~and $\Lambda(\overline{\Lambda})$~candidates are reconstructed through their weak decays within $|y|<0.5$ using the dominant decay channels $\mathrm{K}_{S}^{0} \rightarrow \pi^+ + \pi^-$ (branching ratio 69.2\%) and $\Lambda(\overline{\Lambda}) \rightarrow \mathrm{p}(\overline{\mathrm{p}}) + \pi^{\mp}$ (branching ratio 63.9\%), respectively~\cite{burkertPDG2022}. Combinatorial background is suppressed by applying topological selections~\cite{ALICE:2016fzo}. The decay vertex must be at least 0.5 cm from the primary vertex, and the proper decay length is required to be less than 20 cm for $\mathrm{K}_{S}^{0}$~and 30 cm for $\Lambda(\overline{\Lambda})$. Daughter tracks are identified as $\mathrm{\pi}^{\pm}$~or $\mathrm{p}(\overline{\mathrm{p}})$~
within $3\sigma$ of the expected specific ionization energy loss, $\text{d}E/\text{d}x$, in the TPC.
Additional geometric cuts on the distance of closest approach and pointing angle are optimized separately for $\mathrm{K}_{S}^{0}$~and $\Lambda(\overline{\Lambda})$~candidates to further reduce the background.

\section{Analysis procedure}
\subsection{Flow measurement}
The di-hadron correlation method~\cite{ALICE:2023ulm,ALICE:2015lpx,Abelev:2012ola} is employed to extract the $v_{2}(p_{\rm T})$ of identified particles. Within this context, the trigger particles are identified hadrons selected from the central barrel detectors and reconstructed with the ITS and TPC, while the associated particles are inclusive charged hadrons detected in the forward and backward subdetectors, FMDA and FMDC. The di-hadron two-dimensional (2-D) correlation function is constructed as a function of the transverse momentum $p_{\mathrm{T}}$~of the trigger particles in three event-multiplicity ranges, $15 \leq N_{\rm ch}<25$, $25 \leq N_{\rm ch}<50$, and $N_{\rm ch} \geq 50$, where $N_{\rm ch}$ is the number of charged hadrons reconstructed in the TPC with the kinematic range of $0.2 < p_{\rm T} < 3.0$ GeV/$c$ and $|\eta|< 0.8$, and corrected for tracking efficiency. For $\mathrm{K}_{S}^{0}$~and $\Lambda(\overline{\Lambda})$, the correlation functions in different $p_{\mathrm{T}}$~intervals are obtained by subtracting the correlation function measured in the invariant-mass sideband region, defined as $4\sigma$ to $7\sigma$ away from the mass peak, from the correlation function extracted within $\pm 3\sigma$ of the mass peak. The invariant-mass distribution is fitted using a double-Gaussian for signal and a third-order polynomial for the background~\cite{ALICE:2024vzv,ALICE:2021nvv}.

Flow measurements in p--Pb and pp collisions are substantially influenced by non-flow contamination arising from jets and resonance decays~\cite{ALICE:2019zfl,ALICE:2024vzv}. Consequently, the implementation of a non-flow suppression technique is essential, particularly at higher $p_{\mathrm{T}}$, where the non-flow contributions become larger. As demonstrated in the previous study~\cite{ALICE:2024vzv}, jet contributions near $\Delta\varphi\sim0$ can be effectively suppressed by constructing 2-D correlation functions with a large pseudorapidity gap between trigger particles from the midrapidity region and associated particles from forward and backward regions. 

In addition to the large pseudorapidity gap technique, the template-fit method~\cite{ATLAS:templateFit} provides an effective way to subtract residual non-flow contributions in small collision systems~\cite{Sjostrand:2007gs,ALICE:2024vzv}. This method assumes that the correlation function measured in high-multiplicity events can be described as a superposition of a flow and a non-flow component, where the latter is quantified using low-multiplicity events as a baseline. For this analysis, events with $N_{\rm ch}<15$ are selected as the non-flow baseline for the template fit in both p--Pb and pp collisions. 
The integration of the long‑range di-hadron correlation technique with the template-fit procedure effectively suppresses non-flow contributions and enables reliable extraction of flow signals in small collision systems.

The 2-D correlation function is projected onto the $\Delta\varphi$ axis, and the resulting distribution $Y(\Delta\varphi)$ is fitted using the template-fit procedure, defined as~\cite{ALICE:2024vzv,ALICE:2025bwp}:
\begin{equation}
    Y(\Delta \varphi) = F\,Y^{\rm LM}(\Delta \varphi) + G\left[1+\sum_{n=2}^{4} 2V_{n\Delta}\cos(n\Delta \varphi)\right].
    \label{eq:TF}
\end{equation}
Here, $Y^{\rm LM}(\Delta \varphi)$ denotes the azimuthal-correlation distribution from the low-multiplicity baseline. The parameters $F$, $G$, and $V_{n\Delta}$ are the free parameters in the template fit~\cite{ATLAS:templateFit}. The $v_{n}\{2\}(p_{\rm T})$ coefficients are measured using the three-subevent method~\cite{ALICE:2022cwa}, according to:
\begin{equation}
    v_n\{2\}(p_{\rm T})=\sqrt{\frac{V_{n\Delta}^{\rm TPC-FMDA}(p_{\rm T}) \, V_{n\Delta}^{\rm TPC-FMDC}(p_{\rm T})}{V_{n\Delta}^{\rm FMDA-FMDC}}},
    \label{eq:vn2wFMD}
\end{equation}
where the two-particle correlation $V_{n\Delta}^{\rm FMDA\text{-}FMDC}$ represents the template-fit parameter extracted from the correlation function between particles detected in FMDA and FMDC~\cite{ALICE:2025bwp}. The $p_{\mathrm{T}}$-differential two-particle correlations $V_{n\Delta}^{\rm TPC\text{-}FMDA}$ and $V_{n\Delta}^{\rm TPC\text{-}FMDC}$ are extracted from template fits to the correlation functions between trigger particles reconstructed in the TPC within a given $p_{\mathrm{T}}$~interval and associated particles detected in FMDA and FMDC, respectively.

\subsection{Systematic uncertainties}
\label{sec:syst}
The systematic uncertainties are estimated from the relative differences between the elliptic flow measurements obtained with the default analysis settings and those obtained by varying the event and track-selection criteria, particle-identification procedure, Monte Carlo closure tests, and non-flow baseline modeling, as described below.

For the systematic checks related to variations in event selection, the $z$-position of the primary vertex is tightened from $|\mathrm{PV}_{z}| < 8$~cm to $|\mathrm{PV}_{z}| < 6$~cm, resulting in a 2\% deviation of $v_{2}$ in both p--Pb and pp collisions. The out-of-bunch pileup rejection, determined from the FMD--V0 correlation, is also varied, producing deviations of 1\% in p--Pb and 7\% in pp collisions. Beyond these event-selection variations, track-reconstruction quality criteria are modified to estimate their contributions to the total systematic uncertainty. Specifically, the minimum number of TPC space points is increased from 70 to 90, and the $\chi^{2}_{\mathrm{TPC}}/\mathrm{NDF}$ requirement for the track fit is tightened from 2.5 to 1.0, with each change leading to deviations below 1\% in both collision systems. The impact of secondary-particle contamination on tracks reconstructed within midrapidity is evaluated by tightening the DCA selection. Tightening the $p_{\mathrm{T}}$-dependent $\mathrm{DCA}_{xy}$ selection to $\mathrm{DCA}_{xy} < 5 \times (0.0026 + 0.005/p_{\mathrm{T}}^{1.01})~\mathrm{cm}$ results in deviations of 1\% in p--Pb collisions and 3\% in pp collisions. Tightening the $\mathrm{DCA}_{z}$ requirement from 2~cm to 1~cm yields deviations below 1\%. A Monte Carlo closure test performed on generated and reconstructed samples using GEANT3~\cite{Brun:1994aa} indicates a 4\% systematic uncertainty originating from secondary contamination in the FMD. 
To investigate the systematic effects arising from the FMD material budget, two MC simulations with the budget increased or decreased by 10\% are performed, and the resulting variations of 2\% are assigned as a systematic uncertainty. Systematic uncertainties associated with particle identification are evaluated by tightening the Bayesian PID selection to increase purity, resulting in deviations of $v_{2}$ of $\mathrm{\pi}^{\pm}$, $\mathrm{K}^{\pm}$~and $\mathrm{p}(\overline{\mathrm{p}})$~ of 3.7\% on average and 10\% or more at high $p_{\rm T}$. For $\mathrm{K}_{S}^{0}$~and $\Lambda(\overline{\Lambda})$~candidates, the systematic variations are examined by varying the topological selection criteria, and are found to be below 1\%. The signal extraction for $\mathrm{K}_{S}^{0}$~and $\Lambda(\overline{\Lambda})$~is tested by adjusting the fit function and the fit range of the invariant-mass distributions, also yielding deviations below 1\%. 
The dominant systematic contribution arises from the selection of the non-flow baseline used in the template fit. Varying the multiplicity range used for the baseline from $N_{\rm ch}<15$ to $N_{\rm ch}<10$ leads to the largest systematic uncertainty of 4\% in p--Pb and 7\% in pp collisions. 
The Barlow criterion~\cite{Barlow:2002yb} is used to evaluate whether the variations associated with each systematic check exceed the expected statistical fluctuations, taking into account whether the variation is correlated or uncorrelated with the default measurement. Variations exceeding 1$\sigma$ are included as systematic uncertainty contributions and summed in quadrature to obtain the total systematic uncertainty.

\section{Results}
\begin{figure*}[!ht]
\centering
\begin{minipage}[htbp]{1.0 \linewidth}
    \centering
    \includegraphics[width=1.\linewidth]{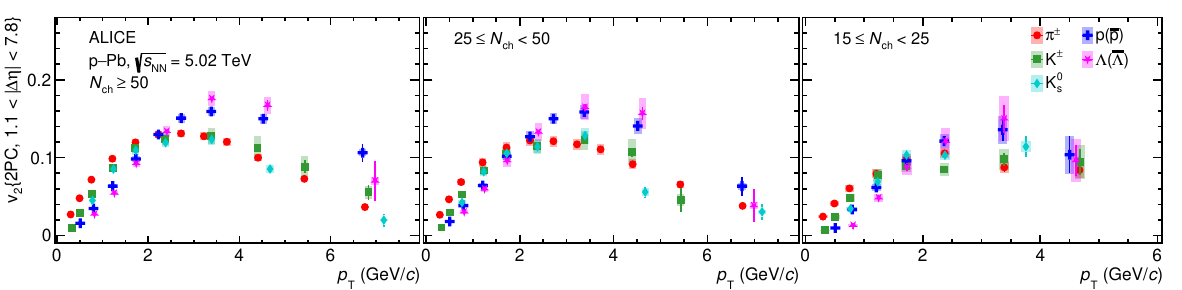}
\end{minipage}
\caption{The $p_{\mathrm{T}}$-differential $v_{2}$ measured through 2-particle correlations (2PC) for identified particles in three multiplicity ranges in p--Pb collisions at $\sqrt{s_{\rm NN}}$~=~5.02~TeV. From left to right, the panels show $N_{\rm ch} \geq 50$, $25 \leq N_{\rm ch}<50$, and $15 \leq N_{\rm ch}<25$. The $p_{\rm T}$ intervals are selected separately for each particle species and multiplicity class. Each point is placed at the mean $p_{\rm T}$ calculated from the corresponding particle $p_{\rm T}$ spectrum within that interval. The vertical error bars and the filled boxes represent statistical and systematic uncertainties, respectively.}
\label{fig:v2pt_pPb}
\end{figure*}

Figure~\ref{fig:v2pt_pPb} presents the measured $p_{\mathrm{T}}$-differential $v_{2}$ for mesons ($\mathrm{\pi}^{\pm}$, $\mathrm{K}^{\pm}$, and $\mathrm{K}_{S}^{0}$) and baryons ($\mathrm{p}(\overline{\mathrm{p}})$~and $\Lambda(\overline{\Lambda})$) in p--Pb collisions at $\sqrt{s_{\rm NN}}$~=~5.02~TeV, obtained across three multiplicity ranges: $N_{\rm ch} \geq 50$, $25 \leq N_{\rm ch}<50$, and $15 \leq N_{\rm ch}<25$. The overall magnitude of $v_{2}$ for all identified particles exhibits a weak dependence on event multiplicity, which is consistent with previous measurements of the multiplicity dependence of flow coefficients~\cite{ALICE:2019zfl}. In every panel of Fig.~\ref{fig:v2pt_pPb}, a clear mass ordering is observed for the low $p_{\mathrm{T}}$~interval ($p_{\rm T}<2$~GeV/$c$) in all multiplicity regions, with the $v_{2}$ of $\mathrm{\pi}^{\pm}$~being the largest and the $v_{2}$ of $\Lambda(\overline{\Lambda})$~being the smallest. The observed mass ordering is interpreted as the combined effect of radial and anisotropic flow. 
Previous studies have shown that theoretical models incorporating either fragmentation of saturated gluon states~\cite{Schenke:2016lrs}, hadronic interactions alone~\cite{Zhou:2015iba,Bierlich:2024odg}, or partonic interactions~\cite{Zhao:2020wcd,YuanyuanWang:2023vor,Tang:2023wcd} can qualitatively reproduce the mass ordering of identified particle flow at low $p_{\mathrm{T}}$. However, a quantitative description of the flow magnitudes for identified hadrons in small collision systems has so far been achieved only within hydrodynamic frameworks~\cite{Zhao:2020wcd,YuanyuanWang:2023vor}, where anisotropic flow develops predominantly during the QGP phase. This agreement therefore suggests the formation of an expanding strongly interacting fluid.
The crossing points of $v_{2}$ for different identified particles are located near $p_{\rm T}\approx2.2$~GeV/$c$ in all multiplicity regions. In the intermediate $p_{\mathrm{T}}$~interval ($2.5<p_{\rm T}<6$~GeV/$c$), the $v_{2}$ values of baryons (mesons) exhibit a characteristic grouping behavior, with deviations within $0.1$–$1.5\sigma$ among species in all multiplicity regions. 
This grouping effect at intermediate $p_{\mathrm{T}}$~is typically reproduced by models incorporating quark-coalescence mechanisms, reflecting the presence of partonic degrees of freedom~\cite{Zhao:2020wcd,YuanyuanWang:2023vor}.
To quantify the splitting, the $v_{2}$ values are averaged over $3<p_{\rm T}<5$~GeV/$c$ for each species and subsequently combined to obtain the mean $v_{2}$ values for baryons and mesons, assuming uncorrelated uncertainties. The baryon--meson splitting significance ($\sigma$) in the intermediate $p_{\mathrm{T}}$~interval is evaluated relative to zero and is defined as the difference between the averaged baryon and meson $v_{2}$ values divided by their combined uncertainties, where statistical and systematic uncertainties are added in quadrature. Baryon--meson splitting is observed in the $N_{\rm ch} \geq 50$ and $25 \leq N_{\rm ch}<50$ ranges, with extracted significances of $6.13\sigma$ and $5.28\sigma$, respectively. In the lowest multiplicity region, $15 \leq N_{\rm ch}<25$, as shown in the right panel of Fig.~\ref{fig:v2pt_pPb}, the splitting is approximately $1.95\sigma$. 

\begin{figure*}[!ht]
\centering
\begin{minipage}[htbp]{0.45\linewidth}
    \centering
    \makebox[\linewidth][c]{\includegraphics[width=1.1\linewidth]{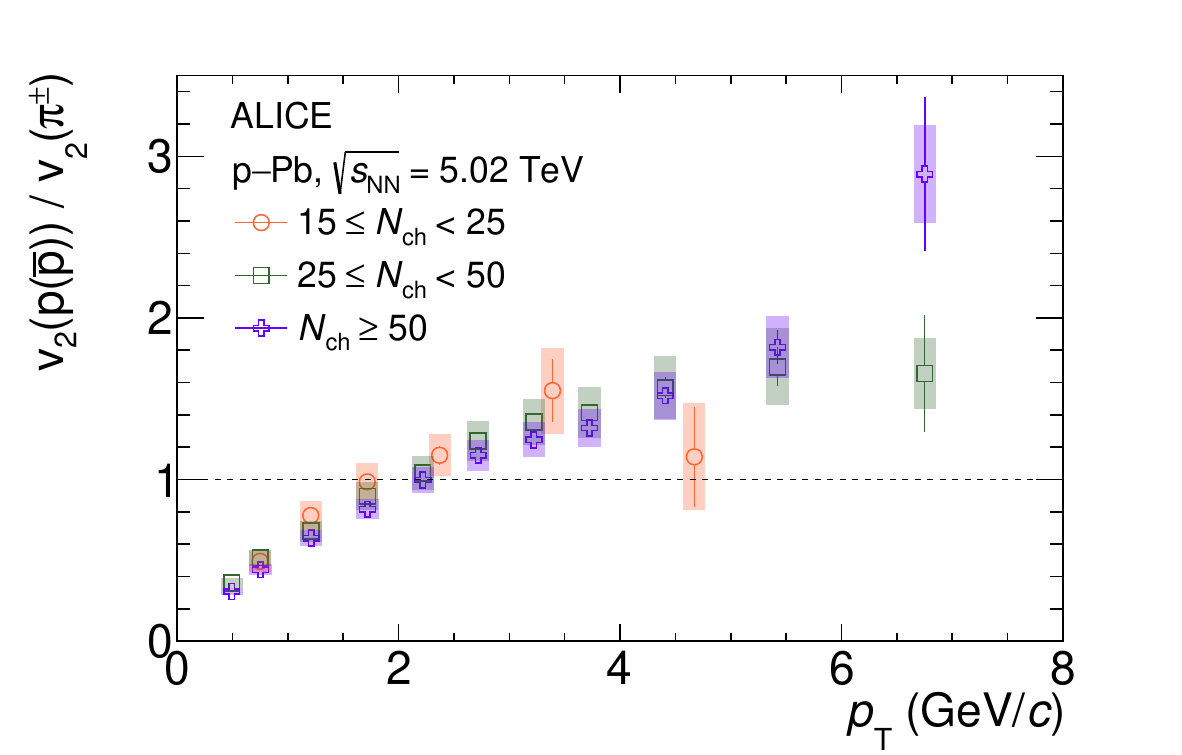}}
\end{minipage}
\begin{minipage}[htbp]{0.45\linewidth}
    \centering
    \makebox[\linewidth][c]{\includegraphics[width=1.1\linewidth]{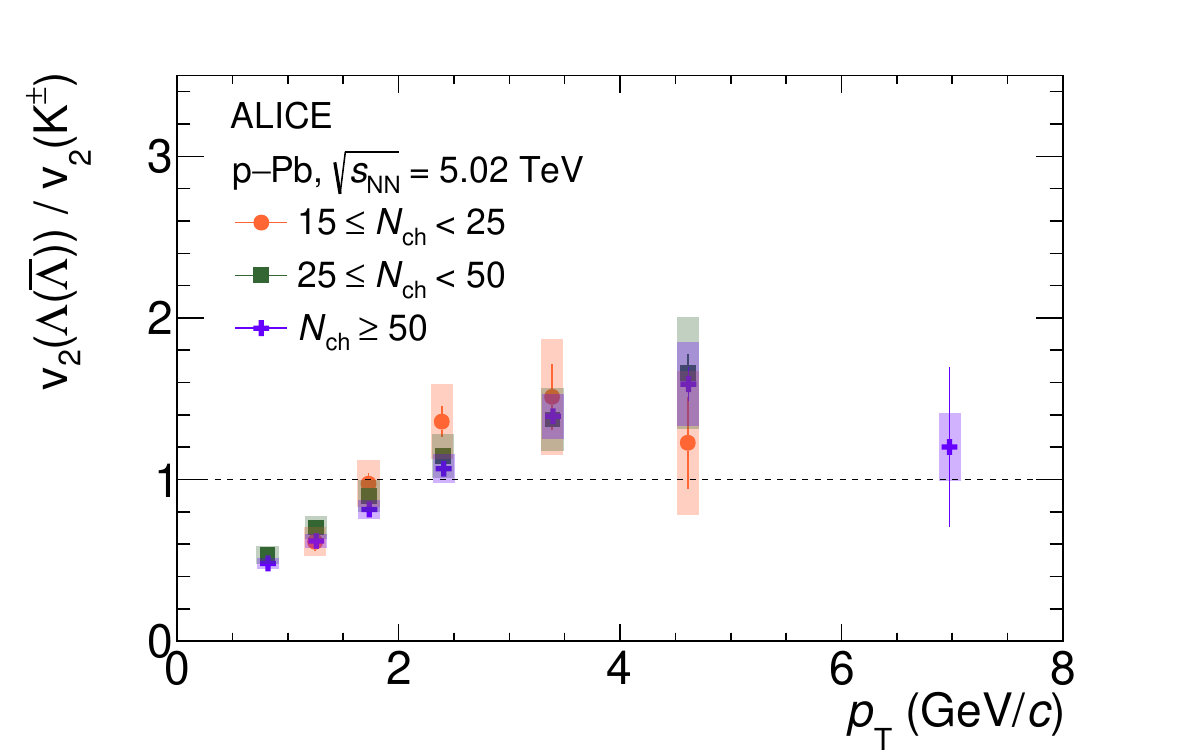}}
\end{minipage}
\caption{The dependence of $v_{2}(\mathrm{p}(\overline{\mathrm{p}}))/v_{2}(\pi^{\pm})$ (left) and $v_{2}(\Lambda(\overline{\Lambda}))/v_{2}(\mathrm{K}^{\pm})$ (right) on transverse momentum for different multiplicity ranges in p--Pb collisions at $\sqrt{s_{\rm NN}}$~=~5.02~TeV. The vertical error bars and the filled boxes represent statistical and systematic uncertainties, respectively.}
\label{fig:v2ra_pPb}
\end{figure*}

The observed baryon--meson grouping and splitting patterns in the $N_{\rm ch} \geq 25$ multiplicity classes in p--Pb collisions are consistent with the presence of partonic collectivity~\cite{Zhao:2020wcd, YuanyuanWang:2023vor}. To quantify the baryon-meson grouping and splitting, Fig.~\ref{fig:v2ra_pPb} shows the baryon‑to‑meson $v_{2}$ ratios as functions of $p_{\mathrm{T}}$~for the three multiplicity classes in p--Pb collisions.
The left panel presents $v_{2}(\mathrm{p}(\overline{\mathrm{p}}))/v_{2}(\pi^{\pm})$, while the right panel presents $v_{2}(\Lambda(\overline{\Lambda}))/v_{2}(\mathrm{K}^{\pm})$. To account for the different $p_{\mathrm{T}}$~intervals among species, spline interpolation is used to obtain the $v_{2}(\pi^{\pm})$($v_{2}(\mathrm{K}^{\pm})$) values at the $p_{\mathrm{T}}$~points of $v_{2}(\mathrm{p}(\overline{\mathrm{p}}))$($v_{2}(\Lambda(\overline{\Lambda}))$).
Both statistical and systematic uncertainties are evaluated through the spline interpolation procedure and subsequently propagated to the resulting ratios, assuming the results in both numerator and denominator are fully independent.
Across all multiplicity classes, the ratios increase from low to intermediate $p_{\mathrm{T}}$, crossing unity near $p_{\rm T} \approx 2.2$~GeV/$c$. For $N_{\rm ch} \geq 50$ and $25 \leq N_{\rm ch}<50$, both $v_{2}(\mathrm{p}(\overline{\mathrm{p}}))/v_{2}(\pi^{\pm})$ and $v_{2}(\Lambda(\overline{\Lambda}))/v_{2}(\mathrm{K}^{\pm})$ continue to rise and remain above unity over a broader  $p_{\mathrm{T}}$~range. In the $15 \leq N_{\rm ch}<25$ class, the ratios slightly increase beyond the crossing point at $p_{\rm T} \approx 2.2$~GeV/$c$ but become consistent with unity at $4<p_{\rm T}<6$ GeV/$c$. No significant deviation from unity in the baryon-to-meson $v_{2}$ ratios at high $p_{\mathrm{T}}$~is observed in this multiplicity class.
Although the current statistical limitations prevent a clear measurement at $p_{\rm T}>8$ GeV/$c$, string-fragmentation mechanisms are generally expected to reduce the difference in anisotropy between produced baryons and mesons, which would lead to baryon-to-meson $v_{2}$ ratios approaching unity~\cite{Zhao:2020wcd,YuanyuanWang:2023vor,Sjostrand:2007gs,Cao:2013ita}.

\begin{figure*}[!ht]
\centering
\begin{minipage}[htbp]{0.95\linewidth}
    \centering
    \includegraphics[width=1.\linewidth]{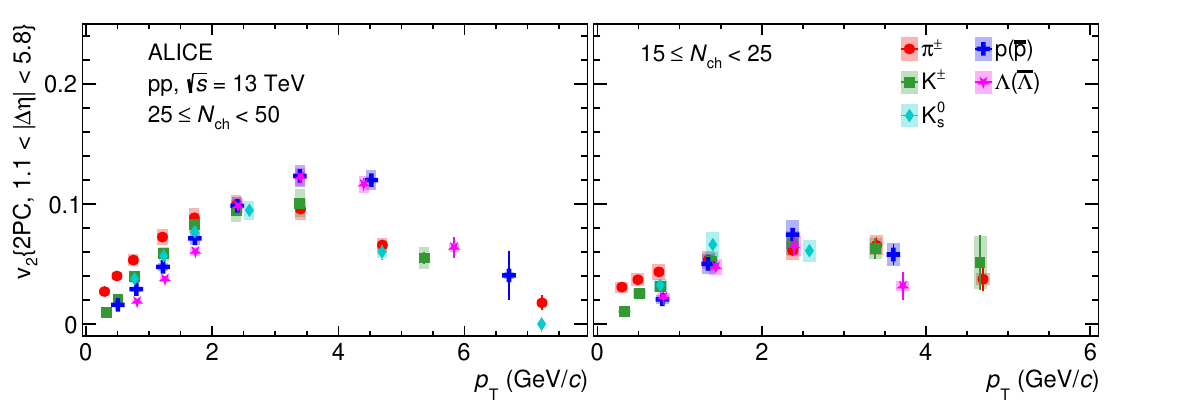}
\end{minipage}
\caption{The $p_{\mathrm{T}}$-differential $v_{2}$ measured through 2-particle correlations (2PC) for identified particles in two multiplicity ranges in pp collisions at $\sqrt{s}$~=~13~TeV. From left to right, the panels show $25 \leq N_{\rm ch}<50$ and $15 \leq N_{\rm ch}<25$. The vertical error bars and the filled boxes represent statistical and systematic uncertainties, respectively.}
\label{fig:v2pt_pp}
\end{figure*}

Similar measurements in pp collisions at $\sqrt{s}$~=~13~TeV  are performed for the $25 \leq N_{\rm ch}<50$ and $15 \leq N_{\rm ch}<25$ multiplicity classes. The results are presented in Fig.~\ref{fig:v2pt_pp}. For the $25 \leq N_{\rm ch}<50$ class, $v_{2}(p_{\rm T})$ demonstrates clear mass ordering at low $p_{\mathrm{T}}$, as well as baryon--meson grouping and splitting at intermediate $p_{\mathrm{T}}$. The deviations in the $v_2$ values among the grouped particles are within $0.2-1.5\sigma$, while the significance of baryon--meson splitting is approximately $4.78\sigma$. In the low-multiplicity region, $15 \leq N_{\rm ch}<25$, all particle species are grouped, and no clear difference in $v_{2}$ is observed at intermediate $p_{\mathrm{T}}$. The significance of the splitting of $v_{2}$ between baryons and mesons is approximately $0.35\sigma$. While the statistical limitations of the current ALICE dataset prevent definitive conclusions, the results suggest no significant baryon--meson splitting signal in low-multiplicity pp collisions.

\begin{figure*}[!ht]
\centering
\begin{minipage}[htbp]{0.45\linewidth}
    \centering
    \makebox[\linewidth][c]{\includegraphics[width=1.1\linewidth]{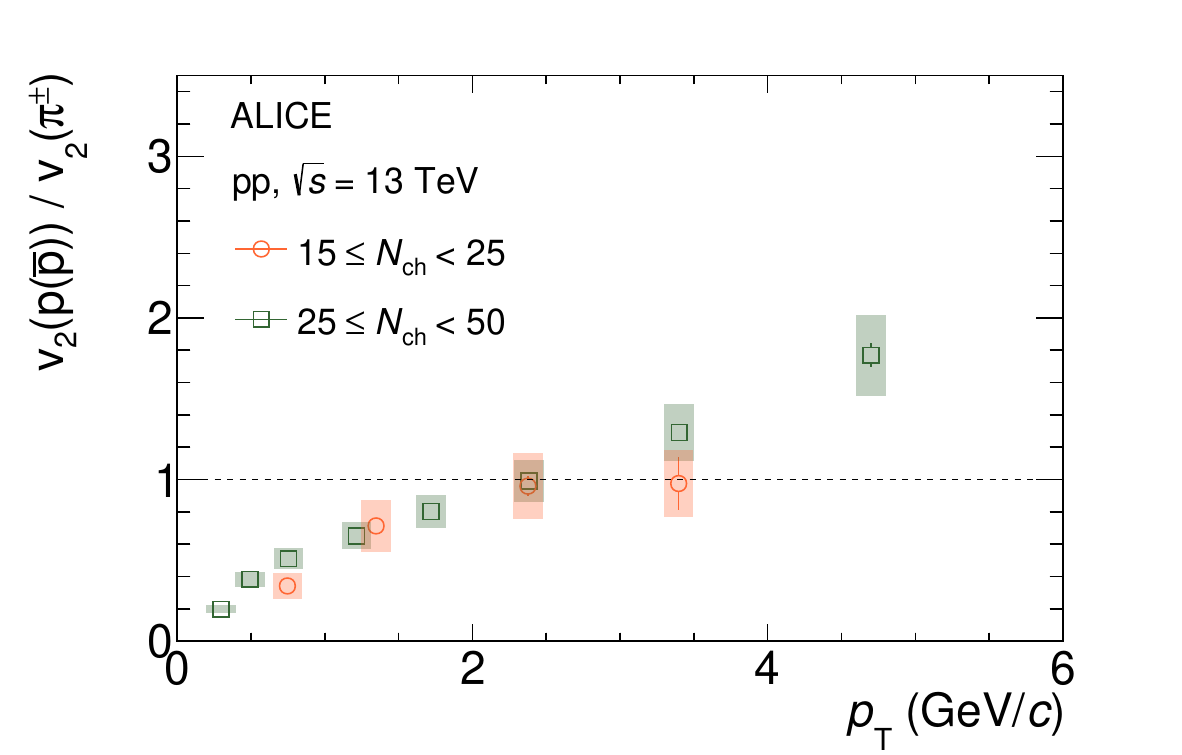}}
\end{minipage}
\begin{minipage}[htbp]{0.45\linewidth}   
    \centering
    \makebox[\linewidth][c]{\includegraphics[width=1.1\linewidth]{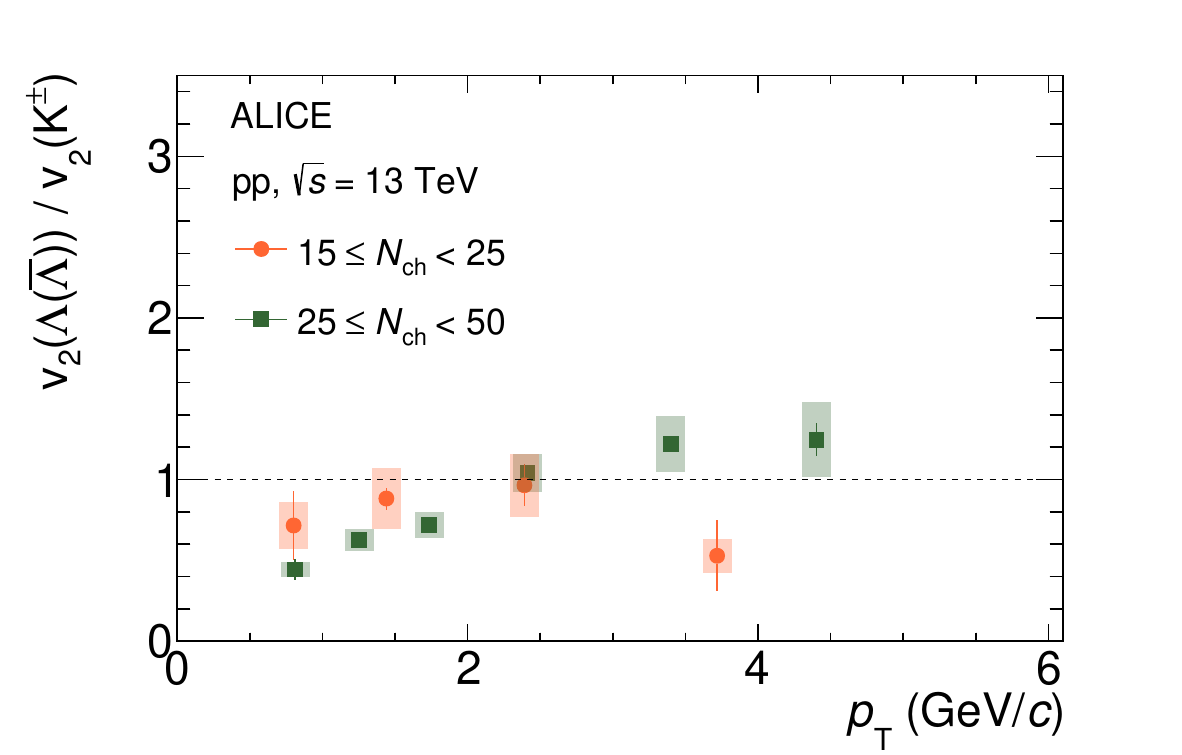}}
\end{minipage}
\caption{The dependence of $v_{2}(\mathrm{p}(\overline{\mathrm{p}}))/v_{2}(\pi^{\pm})$ (left) and $v_{2}(\Lambda(\overline{\Lambda}))/v_{2}(\mathrm{K}^{\pm})$ (right) on transverse momentum for different multiplicity ranges in pp collisions at $\sqrt{s}$~=~13~TeV. The vertical error bars and the filled boxes represent statistical and systematic uncertainties, respectively.}
\label{fig:v2ra_pp}
\end{figure*}

Figure~\ref{fig:v2ra_pp} displays the ratios of $v_{2}$ as a function of $p_{\mathrm{T}}$~for two multiplicity classes in pp collisions. The left panel illustrates the ratio between $\mathrm{p}(\overline{\mathrm{p}})$~and $\mathrm{\pi}^{\pm}$, while the right panel depicts the ratio between $\Lambda(\overline{\Lambda})$~and $\mathrm{K}^{\pm}$. For $25 \leq N_{\rm ch}<50$, both $v_{2}(\mathrm{p}(\overline{\mathrm{p}}))/v_{2}(\pi^{\pm})$ and $v_{2}(\Lambda(\overline{\Lambda}))/v_{2}(\mathrm{K}^{\pm})$ demonstrate trends similar to those in high-multiplicity p--Pb collisions (Fig.~\ref{fig:v2ra_pPb}), including an increasing trend in the low and intermediate $p_{\mathrm{T}}$~intervals, with the ratios crossing unity. In the low-multiplicity region $15 \leq N_{\rm ch}<25$, $v_{2}(\mathrm{p}(\overline{\mathrm{p}}))/v_{2}(\pi^{\pm})$ crosses unity and subsequently remains consistent with unity within uncertainties. In the same multiplicity range, $v_{2}(\Lambda(\overline{\Lambda}))/v_{2}(\mathrm{K}^{\pm})$ stays below unity.

When comparing the $v_{2}$ measurements shown in Figs.~\ref{fig:v2pt_pPb} and~\ref{fig:v2pt_pp}, one finds that the $v_{2}$ values in p--Pb collisions are larger than those in pp collisions, consistent with previous results~\cite{ALICE:2019zfl,ALICE:2024vzv}. It should be noted, however, that the same $N_{\rm ch}$ class in p--Pb and pp collisions does not necessarily correspond to identical underlying physics in the two systems.
Similar qualitative patterns are observed in the measured $v_{2}(p_{\rm T})$ as well as the baryon-to-meson ratios of $v_{2}(p_{\rm T})$ for events with $N_{\rm ch} \geq 25$ in both p--Pb and pp collisions. For the multiplicity class $15 \leq N_{\rm ch}<25$, $v_{2}(p_{\rm T})$ exhibits different behaviors in p--Pb and pp collisions at intermediate $p_{\mathrm{T}}$. Baryon--meson splitting has a significance of approximately $2\sigma$ in p--Pb collisions, and less than $1\sigma$ in pp collisions. The baryon-to-meson $v_{2}$ ratios exceed and return to unity in p--Pb collisions, while the ratios fluctuate around unity in pp collisions.

\section{Conclusions}
This letter presents measurements of $v_{2}(p_{\rm T})$ of identified particles in different multiplicity classes in p--Pb and pp collisions at the LHC. In the high-multiplicity region $N_{\rm ch} \geq 25$, both p--Pb and pp collisions demonstrate a clear mass ordering at $p_{\rm T} < 3$~GeV/$c$, as well as a baryon--meson grouping and splitting at $3 < p_{\rm T} < 6$~GeV/$c$. The baryon-to-meson flow ratios exceed unity over a broad $p_{\mathrm{T}}$~interval, suggesting that quark coalescence has a major contribution to the particle production. 
In p--Pb collisions with $15 \leq N_{\rm ch} < 25$, mass ordering and grouping of $v_{2}$ for identified particles remain visible. The baryon-to-meson $v_{2}$ splitting, as quantified by the baryon-to-meson $v_{2}$ ratios, exhibits a pattern consistent with that observed at higher multiplicities, with the ratios remaining above unity for $p_{\rm T} > 2.2$ GeV/$c$. However, the sizable experimental uncertainties reduce the significance of the splitting to $\sim2\sigma$. In pp collisions within the same multiplicity range, the baryon-to-meson $v_{2}$ ratios are consistent with or below unity, and the significance of the baryon-to-meson splitting is less than $1\sigma$. Given current experimental uncertainties, no firm conclusion can be drawn about the presence or absence of the splitting.
The observed flow across various multiplicity ranges warrants theoretical investigation, particularly in the low-multiplicity regime. High-precision measurements of the elliptic flow of identified particles in the low-multiplicity region, enabled by the recently completed LHC Run~3 data-taking period~\cite{ALICE:2020fuk}, will be crucial to further constrain the limits of partonic collectivity in small collision systems.

\newenvironment{acknowledgement}{\relax}{\relax}
\begin{acknowledgement}
\section*{Acknowledgements}
% Version: 2026-09-09

The ALICE Collaboration would like to thank all its engineers and technicians for their invaluable contributions to the construction of the experiment and the CERN accelerator teams for the outstanding performance of the LHC complex.
The ALICE Collaboration gratefully acknowledges the resources and support provided by all Grid centres and the Worldwide LHC Computing Grid (WLCG) collaboration.
The ALICE Collaboration acknowledges the following funding agencies for their support in building and running the ALICE detector:
A. I. Alikhanyan National Science Laboratory (Yerevan Physics Institute) Foundation (ANSL), State Committee of Science and World Federation of Scientists (WFS), Armenia;
Austrian Academy of Sciences, Austrian Science Fund (FWF): [M 2467-N36] and Nationalstiftung f\"{u}r Forschung, Technologie und Entwicklung, Austria;
Ministry of Communications and High Technologies, National Nuclear Research Center, Azerbaijan;
Rede Nacional de Física de Altas Energias (Renafae), Financiadora de Estudos e Projetos (Finep), Funda\c{c}\~{a}o de Amparo \`{a} Pesquisa do Estado de S\~{a}o Paulo (FAPESP) and The Sao Paulo Research Foundation  (FAPESP), Brazil;
Bulgarian Ministry of Education and Science, within the National Roadmap for Research Infrastructures 2020-2027 (object CERN), Bulgaria;
Ministry of Education of China (MOEC) , Ministry of Science \& Technology of China (MSTC) and National Natural Science Foundation of China (NSFC), China;
Ministry of Science and Education and Croatian Science Foundation, Croatia;
Centro de Aplicaciones Tecnol\'{o}gicas y Desarrollo Nuclear (CEADEN), Cubaenerg\'{\i}a, Cuba;
Ministry of Education, Youth and Sports of the Czech Republic, Czech Republic;
The Danish Council for Independent Research | Natural Sciences, the VILLUM FONDEN and Danish National Research Foundation (DNRF), Denmark;
Helsinki Institute of Physics (HIP), Finland;
Commissariat \`{a} l'Energie Atomique (CEA) and Institut National de Physique Nucl\'{e}aire et de Physique des Particules (IN2P3) and Centre National de la Recherche Scientifique (CNRS), France;
Bundesministerium f\"{u}r Forschung, Technologie und Raumfahrt (BMFTR) and GSI Helmholtzzentrum f\"{u}r Schwerionenforschung GmbH, Germany;
National Research, Development and Innovation Office, Hungary;
Department of Atomic Energy Government of India (DAE), Department of Science and Technology, Government of India (DST), University Grants Commission, Government of India (UGC) and Council of Scientific and Industrial Research (CSIR), India;
National Research and Innovation Agency - BRIN, Indonesia;
Istituto Nazionale di Fisica Nucleare (INFN), Italy;
Japanese Ministry of Education, Culture, Sports, Science and Technology (MEXT) and Japan Society for the Promotion of Science (JSPS) KAKENHI, Japan;
Consejo Nacional de Ciencia (CONACYT) y Tecnolog\'{i}a, through Fondo de Cooperaci\'{o}n Internacional en Ciencia y Tecnolog\'{i}a (FONCICYT) and Direcci\'{o}n General de Asuntos del Personal Academico (DGAPA), Mexico;
Nederlandse Organisatie voor Wetenschappelijk Onderzoek (NWO), Netherlands;
The Research Council of Norway, Norway;
Pontificia Universidad Cat\'{o}lica del Per\'{u}, Peru;
Ministry of Science and Higher Education, National Science Centre and WUT ID-UB, Poland;
Korea Institute of Science and Technology Information and National Research Foundation of Korea (NRF), Republic of Korea;
Ministry of Education and Scientific Research, Institute of Atomic Physics, Ministry of Research and Innovation and Institute of Atomic Physics and Universitatea Nationala de Stiinta si Tehnologie Politehnica Bucuresti, Romania;
Ministerstvo skolstva, vyskumu, vyvoja a mladeze SR, Slovakia;
National Research Foundation of South Africa, South Africa;
Swedish Research Council (VR) and Knut \& Alice Wallenberg Foundation (KAW), Sweden;
European Organization for Nuclear Research, Switzerland;
Suranaree University of Technology (SUT), National Science and Technology Development Agency (NSTDA) and National Science, Research and Innovation Fund (NSRF via PMU-B B05F650021), Thailand;
Turkish Energy, Nuclear and Mineral Research Agency (TENMAK), Turkey;
National Academy of  Sciences of Ukraine, Ukraine;
Science and Technology Facilities Council (STFC), United Kingdom;
National Science Foundation of the United States of America (NSF) and United States Department of Energy, Office of Nuclear Physics (DOE NP), United States of America.
In addition, individual groups or members have received support from:
Czech Science Foundation (grant no. 23-07499S), Czech Republic;
FORTE project, reg.\ no.\ CZ.02.01.01/00/22\_008/0004632, Czech Republic, co-funded by the European Union, Czech Republic;
European Research Council (grant no. 101220549), European Union;
Deutsche Forschungs Gemeinschaft (DFG, German Research Foundation) ``Neutrinos and Dark Matter in Astro- and Particle Physics'' (grant no. SFB 1258), Germany;
CONVECS project, CUP C97H23001700002 FESR 2021-2027 program, Italy.

\end{acknowledgement}

\bibliographystyle{utphys}   \bibliography{ref_v1}

\newpage
\appendix

\section{The ALICE Collaboration}
\label{app:collab}
% ALICE Collaboration author list for 2026-09-09
\begin{flushleft} 
\small

D.A.H.~Abdallah\,\orcidlink{0000-0003-4768-2718}\,$^{\rm 133}$, 
M.M.~Abedin$^{\rm 4}$, 
I.J.~Abualrob\,\orcidlink{0009-0005-3519-5631}\,$^{\rm 110}$, 
S.~Acharya\,\orcidlink{0000-0002-9213-5329}\,$^{\rm 49}$, 
A.B.~Adiguzel$^{\rm 90}$, 
K.~Agarwal\,\orcidlink{0000-0001-5781-3393}\,$^{\rm II,}$$^{\rm 23}$, 
G.~Aglieri Rinella\,\orcidlink{0000-0002-9611-3696}\,$^{\rm 32}$, 
L.~Aglietta\,\orcidlink{0009-0003-0763-6802}\,$^{\rm 24}$, 
N.~Agrawal\,\orcidlink{0000-0003-0348-9836}\,$^{\rm 25}$, 
Z.~Ahammed\,\orcidlink{0000-0001-5241-7412}\,$^{\rm 131}$
% ...

\section*{Affiliation Notes}

$^{\rm I}$ Deceased\\
$^{\rm II}$ Also at: INFN Trieste, Trieste, Italy\\
$^{\rm III}$ Also at: Fondazione Bruno Kessler (FBK), Trento, Italy\\
$^{\rm IV}$ Also at: Czech Technical University in Prague, Prague, Czech Republic\\
$^{\rm V}$ Also at: Dipartimento DET del Politecnico di Torino, Turin, Italy\\
$^{\rm VI}$ Also at: University College of Dublin, Dublin, Ireland\\
$^{\rm VII}$ Also at: Department of Applied Physics, Aligarh Muslim University, Aligarh, India\\
$^{\rm VIII}$ Also at: Institute of Theoretical Physics, University of Wroclaw, Wroclaw, Poland\\
$^{\rm IX}$ Also at: Facultad de Ciencias, Universidad Nacional Aut\'{o}noma de M\'{e}xico, Mexico City, Mexico\\
$^{\rm X}$ Also at: Lovely Professional University, Punjab, India
\\

\section*{Collaboration Institutes}

$^{1}$ A.I. Alikhanyan National Science Laboratory (Yerevan Physics Institute) Foundation, Yerevan, Armenia\\
$^{2}$ AGH University of Krakow, Cracow, Poland\\
$^{3}$ Bogolyubov Institute for Theoretical Physics, National Academy of Sciences of Ukraine, Kyiv, Ukraine\\
$^{4}$ Bose Institute, Department of Physics  and Centre for Astroparticle Physics and Space Science (CAPSS), Kolkata, India\\
$^{5}$ California Polytechnic State University, San Luis Obispo, California, United States\\
$^{6}$ Central China Normal University, Wuhan, China\\
$^{7}$ Centro de Aplicaciones Tecnol\'{o}gicas y Desarrollo Nuclear (CEADEN), Havana, Cuba\\
$^{8}$ Centro de Investigaci\'{o}n y de Estudios Avanzados (CINVESTAV), Mexico City and M\'{e}rida, Mexico\\
$^{9}$ Chicago State University, Chicago, Illinois, United States\\
$^{10}$ China Nuclear Data Center, China Institute of Atomic Energy, Beijing, China\\
$^{11}$ China University of Geosciences, Wuhan, China\\
$^{12}$ Chungbuk National University, Cheongju, Republic of Korea\\
$^{13}$ Comenius University Bratislava, Faculty of Mathematics, Physics and Informatics, Bratislava, Slovak Republic\\
$^{14}$ Creighton University, Omaha, Nebraska, United States\\
$^{15}$ Department of Physics, Aligarh Muslim University, Aligarh, India\\
$^{16}$ Department of Physics, Pusan National University, Pusan, Republic of Korea\\
$^{17}$ Department of Physics, Sejong University, Seoul, Republic of Korea\\
$^{18}$ Department of Physics, University of California, Berkeley, California, United States\\
$^{19}$ Department of Physics, University of Oslo, Oslo, Norway\\
$^{20}$ Department of Physics and Technology, University of Bergen, Bergen, Norway\\
$^{21}$ Dipartimento di Fisica, Universit\`{a} di Pavia, Pavia, Italy\\
$^{22}$ Dipartimento di Fisica dell'Universit\`{a} and Sezione INFN, Cagliari, Italy\\
$^{23}$ Dipartimento di Fisica dell'Universit\`{a} and Sezione INFN, Trieste, Italy\\
$^{24}$ Dipartimento di Fisica dell'Universit\`{a} and Sezione INFN, Turin, Italy\\
$^{25}$ Dipartimento di Fisica e Astronomia dell'Universit\`{a} and Sezione INFN, Bologna, Italy\\
$^{26}$ Dipartimento di Fisica e Astronomia dell'Universit\`{a} and Sezione INFN, Catania, Italy\\
$^{27}$ Dipartimento di Fisica e Astronomia dell'Universit\`{a} and Sezione INFN, Padova, Italy\\
$^{28}$ Dipartimento di Fisica `E.R.~Caianiello' dell'Universit\`{a} and Gruppo Collegato INFN, Salerno, Italy\\
$^{29}$ Dipartimento DISAT del Politecnico and Sezione INFN, Turin, Italy\\
$^{30}$ Dipartimento di Scienze MIFT, Universit\`{a} di Messina, Messina, Italy\\
$^{31}$ Dipartimento Interateneo di Fisica `M.~Merlin' and Sezione INFN, Bari, Italy\\
$^{32}$ European Organization for Nuclear Research (CERN), Geneva, Switzerland\\
$^{33}$ Faculty of Electrical Engineering, Mechanical Engineering and Naval Architecture, University of Split, Split, Croatia\\
$^{34}$ Faculty of Nuclear Sciences and Physical Engineering, Czech Technical University in Prague, Prague, Czech Republic\\
$^{35}$ Faculty of Physics, Sofia University, Sofia, Bulgaria\\
$^{36}$ Faculty of Science, P.J.~\v{S}af\'{a}rik University, Ko\v{s}ice, Slovak Republic\\
$^{37}$ Faculty of Technology, Environmental and Social Sciences, Bergen, Norway\\
$^{38}$ Frankfurt Institute for Advanced Studies, Johann Wolfgang Goethe-Universit\"{a}t Frankfurt, Frankfurt, Germany\\
$^{39}$ Fudan University, Shanghai, China\\
$^{40}$ Gauhati University, Department of Physics, Guwahati, India\\
$^{41}$ Helmholtz-Institut f\"{u}r Strahlen- und Kernphysik, Rheinische Friedrich-Wilhelms-Universit\"{a}t Bonn, Bonn, Germany\\
$^{42}$ Helsinki Institute of Physics (HIP), Helsinki, Finland\\
$^{43}$ High Energy Physics Group,  Universidad Aut\'{o}noma de Puebla, Puebla, Mexico\\
$^{44}$ Horia Hulubei National Institute of Physics and Nuclear Engineering, Bucharest, Romania\\
$^{45}$ HUN-REN Wigner Research Centre for Physics, Budapest, Hungary\\
$^{46}$ Indian Institute of Technology Bombay (IIT), Mumbai, India\\
$^{47}$ Indian Institute of Technology Indore, Indore, India\\
$^{48}$ INFN, Laboratori Nazionali di Frascati, Frascati, Italy\\
$^{49}$ INFN, Sezione di Bari, Bari, Italy\\
$^{50}$ INFN, Sezione di Bologna, Bologna, Italy\\
$^{51}$ INFN, Sezione di Cagliari, Cagliari, Italy\\
$^{52}$ INFN, Sezione di Catania, Catania, Italy\\
$^{53}$ INFN, Sezione di Padova, Padova, Italy\\
$^{54}$ INFN, Sezione di Pavia, Pavia, Italy\\
$^{55}$ INFN, Sezione di Torino, Turin, Italy\\
$^{56}$ INFN, Sezione di Trieste, Trieste, Italy\\
$^{57}$ Inha University, Incheon, Republic of Korea\\
$^{58}$ Institute for Gravitational and Subatomic Physics (GRASP), Utrecht University/Nikhef, Utrecht, Netherlands\\
$^{59}$ Institute of Experimental Physics, Slovak Academy of Sciences, Ko\v{s}ice, Slovak Republic\\
$^{60}$ Institute of Physics of the Czech Academy of Sciences, Prague, Czech Republic\\
$^{61}$ Institute of Space Science (ISS), Bucharest, Romania\\
$^{62}$ Institut f\"{u}r Kernphysik, Johann Wolfgang Goethe-Universit\"{a}t Frankfurt, Frankfurt, Germany\\
$^{63}$ Instituto de Ciencias Nucleares, Universidad Nacional Aut\'{o}noma de M\'{e}xico, Mexico City, Mexico\\
$^{64}$ Instituto de F\'{i}sica, Universidade Federal do Rio Grande do Sul (UFRGS), Porto Alegre, Brazil\\
$^{65}$ Instituto de F\'{\i}sica, Universidad Nacional Aut\'{o}noma de M\'{e}xico, Mexico City, Mexico\\
$^{66}$ iThemba LABS, National Research Foundation, Somerset West, South Africa\\
$^{67}$ Jeonbuk National University, Jeonju, Republic of Korea\\
$^{68}$ Korea Institute of Science and Technology Information, Daejeon, Republic of Korea\\
$^{69}$ Laboratoire de Physique Subatomique et de Cosmologie, Universit\'{e} Grenoble-Alpes, CNRS-IN2P3, Grenoble, France\\
$^{70}$ Lawrence Berkeley National Laboratory, Berkeley, California, United States\\
$^{71}$ Lund University Department of Physics, Division of Particle Physics, Lund, Sweden\\
$^{72}$ Marietta Blau Institute, Vienna, Austria\\
$^{73}$ Nagasaki Institute of Applied Science, Nagasaki, Japan\\
$^{74}$ Nara Women{'}s University (NWU), Nara, Japan\\
$^{75}$ National Centre for Nuclear Research, Warsaw, Poland\\
$^{76}$ National Institute of Science Education and Research, Homi Bhabha National Institute, Jatni, India\\
$^{77}$ National Nuclear Research Center, Baku, Azerbaijan\\
$^{78}$ National Research and Innovation Agency - BRIN, Jakarta, Indonesia\\
$^{79}$ Niels Bohr Institute, University of Copenhagen, Copenhagen, Denmark\\
$^{80}$ Nikhef, National institute for subatomic physics, Amsterdam, Netherlands\\
$^{81}$ Nuclear Physics Group, STFC Daresbury Laboratory, Daresbury, United Kingdom\\
$^{82}$ Nuclear Physics Institute of the Czech Academy of Sciences, Husinec-\v{R}e\v{z}, Czech Republic\\
$^{83}$ Oak Ridge National Laboratory, Oak Ridge, Tennessee, United States\\
$^{84}$ Ohio State University, Columbus, Ohio, United States\\
$^{85}$ Physics Department, Panjab University, Chandigarh, India\\
$^{86}$ Physics Department, University of Jammu, Jammu, India\\
$^{87}$ Physics Program and International Institute for Sustainability with Knotted Chiral Meta Matter (WPI-SKCM$^{2}$), Hiroshima University, Hiroshima, Japan\\
$^{88}$ Physikalisches Institut, Eberhard-Karls-Universit\"{a}t T\"{u}bingen, T\"{u}bingen, Germany\\
$^{89}$ Physikalisches Institut, Ruprecht-Karls-Universit\"{a}t Heidelberg, Heidelberg, Germany\\
$^{90}$ Physik Department, Technische Universit\"{a}t M\"{u}nchen, Munich, Germany\\
$^{91}$ Politecnico di Bari and Sezione INFN, Bari, Italy\\
$^{92}$ Research Division and ExtreMe Matter Institute EMMI, GSI Helmholtzzentrum f\"ur Schwerionenforschung GmbH, Darmstadt, Germany\\
$^{93}$ Saga University, Saga, Japan\\
$^{94}$ Saha Institute of Nuclear Physics, Homi Bhabha National Institute, Kolkata, India\\
$^{95}$ School of Physics and Astronomy, University of Birmingham, Birmingham, United Kingdom\\
$^{96}$ Secci\'{o}n F\'{\i}sica, Departamento de Ciencias, Pontificia Universidad Cat\'{o}lica del Per\'{u}, Lima, Peru\\
$^{97}$ SUBATECH, IMT Atlantique, Nantes Universit\'{e}, CNRS-IN2P3, Nantes, France\\
$^{98}$ Sungkyunkwan University, Suwon City, Republic of Korea\\
$^{99}$ Suranaree University of Technology, Nakhon Ratchasima, Thailand\\
$^{100}$ Technical University of Ko\v{s}ice, Ko\v{s}ice, Slovak Republic\\
$^{101}$ The Henryk Niewodniczanski Institute of Nuclear Physics, Polish Academy of Sciences, Cracow, Poland\\
$^{102}$ The University of Texas at Austin, Austin, Texas, United States\\
$^{103}$ Universidad Aut\'{o}noma de Sinaloa, Culiac\'{a}n, Mexico\\
$^{104}$ Universidade de S\~{a}o Paulo (USP), S\~{a}o Paulo, Brazil\\
$^{105}$ Universidade Estadual de Campinas (UNICAMP), Campinas, Brazil\\
$^{106}$ Universidade Federal do ABC, Santo Andre, Brazil\\
$^{107}$ Universitatea Nationala de Stiinta si Tehnologie Politehnica Bucuresti, Bucharest, Romania\\
$^{108}$ University of Cape Town, Cape Town, South Africa\\
$^{109}$ University of Derby, Derby, United Kingdom\\
$^{110}$ University of Houston, Houston, Texas, United States\\
$^{111}$ University of Jyv\"{a}skyl\"{a}, Jyv\"{a}skyl\"{a}, Finland\\
$^{112}$ University of Kansas, Lawrence, Kansas, United States\\
$^{113}$ University of Liverpool, Liverpool, United Kingdom\\
$^{114}$ University of Science and Technology of China, Hefei, China\\
$^{115}$ University of Silesia in Katowice, Katowice, Poland\\
$^{116}$ University of South-Eastern Norway, Kongsberg, Norway\\
$^{117}$ University of Tennessee, Knoxville, Tennessee, United States\\
$^{118}$ University of the Witwatersrand, Johannesburg, South Africa\\
$^{119}$ University of Tokyo, Tokyo, Japan\\
$^{120}$ University of Tsukuba, Tsukuba, Japan\\
$^{121}$ University of Zagreb Faculty of Science, Department of Physics, Zagreb, Croatia\\
$^{122}$ Universit\"{a}t M\"{u}nster, Institut f\"{u}r Kernphysik, M\"{u}nster, Germany\\
$^{123}$ Universit\'{e} Clermont Auvergne, CNRS/IN2P3, LPC, Clermont-Ferrand, France\\
$^{124}$ Universit\'{e} de Lyon, CNRS/IN2P3, Institut de Physique des 2 Infinis de Lyon, Lyon, France\\
$^{125}$ Universit\'{e} de Strasbourg, CNRS, IPHC UMR 7178, F-67000 Strasbourg, France, Strasbourg, France\\
$^{126}$ Universit\'{e} Paris-Saclay, Centre d'Etudes de Saclay (CEA), IRFU, D\'{e}partment de Physique Nucl\'{e}aire (DPhN), Saclay, France\\
$^{127}$ Universit\'{e}  Paris-Saclay, CNRS/IN2P3, IJCLab, Orsay, France\\
$^{128}$ Universit\`{a} degli Studi di Foggia, Foggia, Italy\\
$^{129}$ Universit\`{a} del Piemonte Orientale, Vercelli, Italy\\
$^{130}$ Universit\`{a} di Brescia, Brescia, Italy\\
$^{131}$ Variable Energy Cyclotron Centre, Homi Bhabha National Institute, Kolkata, India\\
$^{132}$ Warsaw University of Technology, Warsaw, Poland\\
$^{133}$ Wayne State University, Detroit, Michigan, United States\\
$^{134}$ Yale University, New Haven, Connecticut, United States\\
$^{135}$ Yildiz Technical University, Istanbul, Turkey\\
$^{136}$ Yonsei University, Seoul, Republic of Korea\\
$^{137}$ Affiliated with an institute formerly covered by a cooperation agreement with CERN\\
$^{138}$ Affiliated with an international laboratory covered by a cooperation agreement with CERN.\\

\end{flushleft} 

\end{document}